\documentclass[manuscript]{acmart}
\AtBeginDocument{%
  \providecommand\BibTeX{{%
    \normalfont B\kern-0.5em{\scshape i\kern-0.25em b}\kern-0.8em\TeX}}}

\copyrightyear{2024}
\acmYear{2024}
\setcopyright{rightsretained}
\acmConference[ASSETS '24]{The 26th International ACM SIGACCESS Conference on Computers and Accessibility}{October 27--30, 2024}{St. John's, NL, Canada}
\acmBooktitle{The 26th International ACM SIGACCESS Conference on Computers and Accessibility (ASSETS '24), October 27--30, 2024, St. John's, NL, Canada}
\acmDOI{10.1145/3663548.3688502}
\acmISBN{979-8-4007-0677-6/24/10}

\usepackage{multirow}
\usepackage{subfig}
\usepackage{subcaption}
\begin{document}

%%
%% The "title" command has an optional parameter,
%% allowing the author to define a "short title" to be used in page headers.
\title[AI Supported Just-in-Time Programming]{Exploring the use of Generative AI to Support Automated Just-in-Time Programming for Visual Scene Displays}

%%
%% The "author" command and its associated commands are used to define
%% the authors and their affiliations.
%% Of note is the shared affiliation of the first two authors, and the
%% "authornote" and "authornotemark" commands
%% used to denote shared contribution to the research.
%\author{Anonymous Author(s)}
\author{Cynthia Zastudil}
\email{cynthia.zastudil@temple.edu}
\affiliation{%
 \institution{Temple University}
 \streetaddress{1801 N. Broad St.}
 \city{Philadelphia}
 \state{Pennsylvania}
 \country{USA}
 \postcode{19122}
}

\author{Christine Holyfield}
\email{ceholfi@uark.edu}
\affiliation{%
\institution{University of Arkansas}
\streetaddress{1 University of Arkansas}
\city{Fayetteville}
\state{Arkansas}
\country{USA}
\postcode{72701}}

\author{Christine Kapp}
\email{christine.kapp@temple.edu}
\affiliation{%
 \institution{Temple University}
 \streetaddress{1801 N. Broad St.}
 \city{Philadelphia}
 \state{Pennsylvania}
 \country{USA}
 \postcode{19122}
}

\author{Xandria Crosland}
\email{xcrosl1@wgu.edu}
\affiliation{%
 \institution{Western Governors University}
 \streetaddress{4001 S 700 E \#300}
 \city{Millcreek}
 \state{Utah}
 \country{USA}
 \postcode{84107}
}

\author{Elizabeth Lorah}
\email{lorah@uark.edu}
\affiliation{%
\institution{University of Arkansas}
\streetaddress{1 University of Arkansas}
\city{Fayetteville}
\state{Arkansas}
\country{USA}
\postcode{72701}}

\author{Tara Zimmerman}
\email{taraz@uark.edu}
\affiliation{%
\institution{University of Arkansas}
\streetaddress{1 University of Arkansas}
\city{Fayetteville}
\state{Arkansas}
\country{USA}
\postcode{72701}}

\author{Stephen MacNeil}
\email{stephen.macneil@temple.edu}
\affiliation{%
 \institution{Temple University}
 \streetaddress{1801 N. Broad St.}
 \city{Philadelphia}
 \state{Pennsylvania}
 \country{USA}
 \postcode{19122}
}

%%
%% By default, the full list of authors will be used in the page
%% headers. Often, this list is too long, and will overlap
%% other information printed in the page headers. This command allows
%% the author to define a more concise list
%% of authors' names for this purpose.
\renewcommand{\shortauthors}{Zastudil et al.}

%%
%% The abstract is a short summary of the work to be presented in the
%% article.
\begin{abstract}

Millions of people worldwide rely on alternative and augmentative communication devices to communicate.  Visual scene displays (VSDs) can enhance communication for these individuals by embedding communication options within contextualized images. However, existing VSDs often present default images that may lack relevance or require manual configuration, placing a significant burden on communication partners. In this study, we assess the feasibility of leveraging large multimodal models (LMM), such as GPT-4V, to automatically create communication options for VSDs. Communication options were sourced from a LMM and speech-language pathologists (SLPs) and AAC researchers (N=13) for evaluation through an expert assessment conducted by the SLPs and AAC researchers. We present the study's findings, supplemented by insights from semi-structured interviews (N=5) about SLP's and AAC researchers' opinions on the use of generative AI in agumentative and alternative communication devices. Our results indicate that the communication options generated by the LMM were contextually relevant and often resembled those created by humans. However, vital questions remain that must be addressed before LMMs can be confidently implemented in AAC devices.

%aimed at eliciting SLPs' and VSD researchers’ opinions on the automated recognition of objects within photographs used in VSDs and on the ethics of using generative AI in augmentative and alternative communication devices. 
\end{abstract}

%%
%% The code below is generated by the tool at http://dl.acm.org/ccs.cfm.
%% Please copy and paste the code instead of the example below.
%%
\begin{CCSXML}
<ccs2012>
   <concept>
       <concept_id>10003120.10011738.10011775</concept_id>
       <concept_desc>Human-centered computing~Accessibility technologies</concept_desc>
       <concept_significance>500</concept_significance>
       </concept>
   <concept>
       <concept_id>10003120.10011738.10011776</concept_id>
       <concept_desc>Human-centered computing~Accessibility systems and tools</concept_desc>
       <concept_significance>500</concept_significance>
       </concept>
 </ccs2012>
\end{CCSXML}

\ccsdesc[500]{Human-centered computing~Accessibility technologies}
\ccsdesc[500]{Human-centered computing~Accessibility systems and tools}

%%
%% Keywords. The author(s) should pick words that accurately describe
%% the work being presented. Separate the keywords with commas.
\keywords{AAC, autism, visual screen displays, VSDs, generative AI, just-in-time programming}

%% A "teaser" image appears between the author and affiliation
%% information and the body of the document, and typically spans the
%% page.
%\begin{teaserfigure}
%  \includegraphics[width=\textwidth]{sampleteaser}
%  \caption{Seattle Mariners at Spring Training, 2010.}
%  \Description{Enjoying the baseball game from the third-base
%  seats. Ichiro Suzuki preparing to bat.}
%  \label{fig:teaser}
%\end{teaserfigure}

\received{3 July 2024}
\received[Accepted]{9 August 2024}

%%
%% This command processes the author and affiliation and title
%% information and builds the first part of the formatted document.
\maketitle

\section{Introduction \& Background}

%In the United States alone, an estimated 2 million people rely on augmentative and alternative communication (AAC) devices to communicate~\cite{beukelman2020augmentative}. Furthermore, there are millions of children in the United States~\cite{shriberg_prevalence_1999, law_prevalence_2000} who are on the autism spectrum\footnote{Researchers (and everyone) should honor individuals' preferences for how they are identified. While recently many adults on the autism spectrum report preferring identity-first language (e.g., "autistic adult"), this preference is not universal and parents of children may prefer individual-first language (e.g., "child with autism"). We are using the terminology "on the autism spectrum" as it is considered a compromise that is not likely to offend people who prefer individual-first identification (because it follows the initial reference to the person such as "child" or "person") or people who prefer identity-first identification (because it does not separate the person from the diagnoses)~\cite{bury2023defines, dwyer2022first}.} or have other developmental disabilities that do not have functional speech and would therefore benefit from AAC. For these children, early intervention with AAC can be very beneficial for language acquisition and the development of typical verbal language skills~\cite{branson_use_2009, woods_early_2003}. 

\begin{figure*}
\centering
\includegraphics[width=0.7\linewidth]{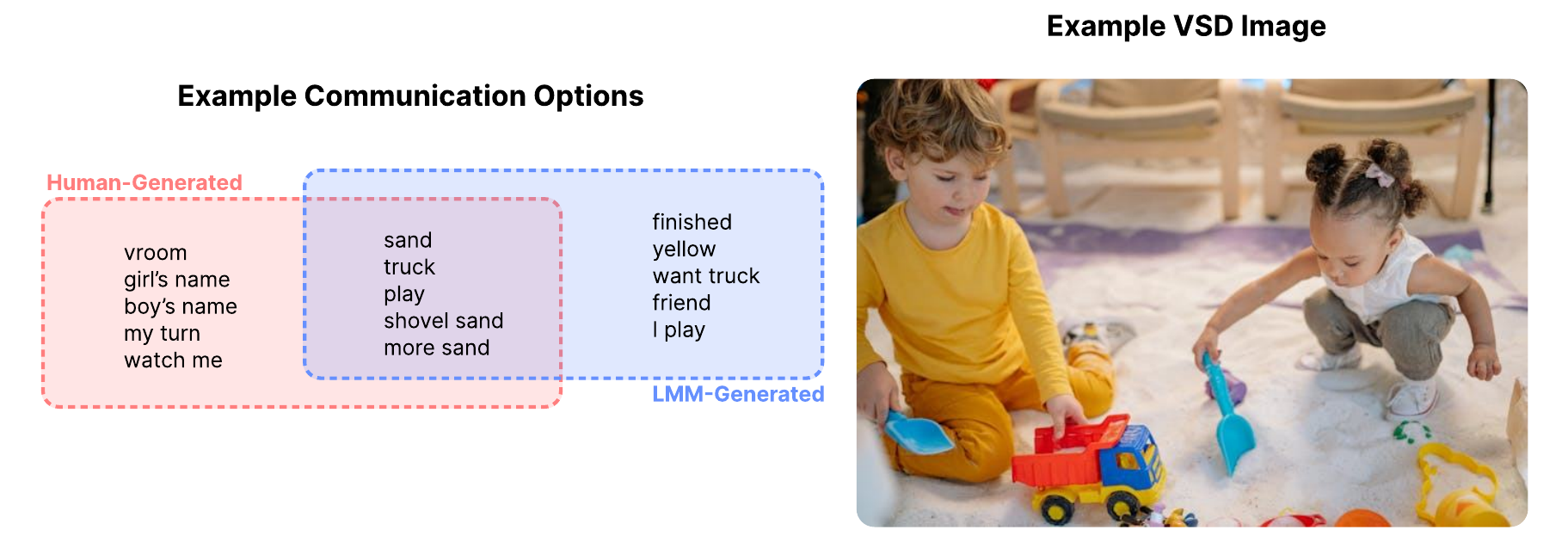}
\caption{An example image which could be used in a VSD. Example COs generated by human participants and by the LMM are provided in the figure. The COs can be embedded within the image as clickable ``hotspots'' or as buttons presented on the display.}
\label{fig:sample-vsd-w-comm-options}
\Description[Two children playing with sand toys and a toy truck]{On the right, is an image which would be used within a visual scene display. Two children are sitting in a sandbox playing with sand toys, specifically blue sand shovels, and a toy dump truck that is red, blue, and yellow. One child appears to be filling up the dump truck with sand using their shovel. The other child appears to be digging in the sand. On the left, example communication options generated by humans and by a large multimodal model (LMM) are presented in a venn diagram format. The communication options generated only by humans are: vroom, girl's name, boy's name, my turn, and watch me. The communication options generated on by the LMM are: finished, yellow, want truck, friend, and I play. The communication options generated by both humans and the LMM are: sand, truck, play, shovel sand, and more sand.}
\end{figure*}

Visual scene displays (VSDs) are a form of augmentative and alternative communication (AAC) which use photographs or other images with interactive ``hotspots'' placed on them to represent language concepts~\cite{blackstone2004visual, wilkinson2011preliminary, light_designing_2019} (see the example in Figure~\ref{fig:sample-vsd-w-comm-options}). VSDs have proven especially useful for beginning communicators (i.e., communicators who are learning their first words) because they incorporate personally relevant imagery~\cite{light2012supporting, wilkinson2011preliminary}, maintain the relationship between people and objects~\cite{light_performance_2004}, combine subjects or activities within a single visual context~\cite{light2012supporting}, and reduce the visual cognitive demands typically associated with AAC by aligning with natural visual processing~\cite{light_designing_2019}.
%These benefits distinguish VSDs from other AAC layouts including the more common grid displays that display symbols, which are often line drawings that do not as clearly reflect referents (i.e., the object, concept, or activity to which the symbol refers), in isolation without maintaining the relationship between words~\cite{mirenda_comparison_1989}.

While VSDs offer numerous benefits for beginning communicators, several challenges can reduce their effectiveness, impacting both their adoption and retention. One significant issue is that the default imagery and communication options (COs) provided by VSDs are only relevant in specific settings, which limits their effectiveness~\cite{smith_asymmetry_2003}. 
%One option to mitigate this issue is manual configuration; however, this requires frequent, time-consuming updates to remain relevant to users\cite{drager_aac_2019}. 
%This manual configuration requires frequent, time-consuming updates to remain relevant for their users’ communication needs~\cite{drager_aac_2019}.
%When VSDs are not updated quickly enough to meet the communicator’s language needs, it can cause the communicator to not be able to actively participate in communication, which can hinder their language development~\cite{clarke_examination_2012}.
%Another more promising option is 
This issue has led to the development of just-in-time (JIT) programming~\cite{drager_aac_2019, holyfield_effect_2019, schlosser2016just, holyfield_programming_2019}, where
%, which we refer to as “just-in-time configuration” throughout this paper to avoid confusion with software programming. 
%JIT programming allows 
communication partners manually create COs for VSD users in real-time, tailored to the specific image or naturally occurring scene.
%It also supports taking advantage of teachable moments as they occur~\cite{schlosser2016just} due to their ability to easily import images and program communication options within VSDs~\cite{drager_aac_2019}.
%Implementing JIT configuration within VSDs has proven to significantly increase communication frequency among children and adolescents~\cite{holyfield_effect_2019}. 
While JIT programming can improve communication outcomes for users~\cite{holyfield_effect_2019},
%there are still concerns as to whether clinicians can adjust accordingly when faced with environment changes and unforeseen situations~\cite{schlosser2016just}. 
it requires clinicians to be present and continuously reconfiguring the user interface to capture contextually relevant and engaging scenarios~\cite{holyfield_effect_2019}. 
%For other communication partners, such as parents or teachers, VSDs with JIT support typically require an upfront appointment or multiple training sessions with a clinician with expertise in AAC to discuss engagement scenarios and relevant communication options~\cite{caron2016operational}.
%This manual configuration approach is supported by evidence~\cite{drager_aac_2019, holyfield_effect_2019, o2016brief, light2012effects}, but requires substantial effort from communication partners~\cite{caron2016operational, caron2017comparison}. 

%Researchers have also explored how information about a user’s context can be used to automatically configure VSDs and other forms of AAC without direct intervention during communication~\cite{devargas2022automated, obiorah2021designing, Mooney_Bedrick_Noethe_Spaulding_Fried-Oken_2018}. Most recently, de Vargas et al. developed a mobile application with a hybrid VSD and grid display that automatically creates communication options and grid displays based on a photograph~\cite{devargas2022automated, devargas2024codesigning}. 
%Researchers have also explored how automated approaches incorporating a user's context (e.g., location or photographs) to automatically configure AAC devices JIT to reduce the effort required by communicators and communication partners~\cite{devargas2022automated, devargas2024codesigning, obiorah2021designing, Mooney_Bedrick_Noethe_Spaulding_Fried-Oken_2018}. 
%Automated approaches may reduce the effort required by communicators and communication partners to maintain contextually relevant VSDs~\cite{devargas2022automated}. 

Prior work has evaluated the potential to automatically generate COs for users. For example, Holyfield et al. investigated the potential to augment a grid display using communication partner speech input, increasing participation of young children using the device~\cite{holyfield2024leveraging}. Prior work has also investigated incorporating automatically generated COs based on a photograph into an AAC device by creating topic-specific grid displays~\cite{devargas2022automated, devargas2024codesigning}.
%There has been preliminary research which found that augmenting a topic-specific grid display with communication partner speech input and color photographs may increase communicative participation of young children on the autism spectrum using this device~\cite{holyfield2024leveraging}. Prior work has also established the feasibility of incorporating automatically generated COs based on a photograph into an AAC device by creating topic-specific grid displays~\cite{devargas2022automated, devargas2024codesigning}.
%With the advent of highly performant large multimodal models (LMMs) (e.g., GPT-4V
%\footnote{\href{https://openai.com/gpt-4}{https://openai.com/gpt-4}},Gemini\footnote{\href{https://gemini.google.com/}{https://gemini.google.com/}}), 
However, it is crucial to carefully examine the content, focus, and relevance of AI-generated communication options (COs), particularly considering the potential for racial and gender biases inherent in AI systems~\cite{navigli2023biases, armstrong2024silicone, ding2023fluid, kotek2023gender}. This raises questions about whether AI-generated communication options can serve as a useful support for beginning communicators.  %Automated COs may be contextually relevant for a given image; however, these automated approaches are typically limited to that context and seldom integrate contextualized information about the end-user themselves. This raises questions about whether options generated from an image can serve as a useful default for beginning communicators.

In this work, we used a LMM to generate COs for VSDs intended for use by young children on the autism spectrum 
%\footnote{Researchers (and everyone) should honor individuals' preferences for how they are identified. While recently many adults on the autism spectrum report preferring identity-first language (e.g., "autistic adult"), this preference is not universal and parents of children may prefer individual-first language (e.g., "child with autism"). We are using the terminology "on the autism spectrum" as it is considered a compromise that is not likely to offend people who prefer individual-first identification (because it follows the initial reference to the person such as "child" or "person") or people who prefer identity-first identification (because it does not separate the person from the diagnosis)~\cite{bury2023defines, dwyer2022first}.}
or with other developmental disabilities. We compared these COs to options created by speech-language pathologists (SLP) and AAC researchers (N=13) to compare how the relevance and focus of the COs differ between the two sets of COs. Expert VSD researchers (N=5) then conducted an evaluation to compare the quality of human- and LMM-generated COs. Lastly, we conducted semi-structured interviews with expert VSD researchers (N=5) to better understand the implications of these models on AAC devices.

%Through these interviews, we uncovered perceptions on the use of generative AI in this context and about the ethical implications of generative AI on AAC technology.
%We focus on VSDs in this work instead of the hybrid display of a photograph with communication options and a grid display proposed by de Vargas et al.~\cite{devargas2022automated} because it would be too complex for young children who are working on emerging language development~\cite{light_designing_2019}. 
%The generative AI models process both image and text data in the same embedding space to understand and produce text and images. 
%We then compared the communication options created by generative AI to options created by speech-language pathologists (SLP) and AAC researchers (N=13) for three different contexts to compare how the relevance and focus of the communication options differs between those generated by humans and generative AI. These communication options were then evaluated using an expert evaluation with SLPs and VSD researchers (N=5). To better understand the implications of these models, we also conducted semi-structured interviews with experts in the fields of speech-language pathology and VSD research. Through these interviews we uncovered perceptions of the use of generative AI in this context as well as insights about the ethical implications of generative AI on AAC technology. 
In this work, we investigate the following research questions: 
%\begin{itemize} 
%\item [\textbf{RQ1}] How do communication options generated by SLPs, AAC researchers, and a LMM compare in terms of perceived relevance, topics of focus, and quality?
%\item [\textbf{RQ2}] What are the perceptions of SLPs and researchers who use VSDs on the use and ethics of the use of LMMs for just-in-time programming of VSDs?
%\end{itemize}
\textbf{(RQ1)} How do communication options generated by SLPs, AAC researchers, and a LMM compare in terms of perceived relevance, topics of focus, and quality? and \textbf{(RQ2)} What are the perceptions of SLPs and researchers who use VSDs on the use and ethics of the use of LMMs for just-in-time programming of VSDs? %Our findings indicate that the content, focus, and quality of LMM-generated COs were generally comparable to those created by humans. However, human-generated COs often included more sound effects and language aimed at social engagement. Furthermore, human-generated COs provided additional contextual information about users that went beyond what was present in any given image. For example, SLPs wanted to include the names of the kids in the photographs. 
% \begin{enumerate}
%     \item [\textbf{RQ1:}] How do communication options generated by SLPs and AAC researchers and a LMM compare in terms of perceived relevance, topics of focus, and quality?
%     %\item [\textbf{RQ 2:}] What preferences do SLPs and researchers who use VSDs have in relation to human-created communication options and AI generated communication options?
%     %\item [\textbf{RQ 2:}] How do human- and LMM-generated communication compare in terms of quality as perceived by SLPs and VSD researchers?
%     \item [\textbf{RQ2:}] What are the perceptions of SLPs and researchers who use VSDs on the use and ethics of the use of LMMs for just-in-time programming of VSDs?
% \end{enumerate}

%Through this research, we have identified significant opportunities and challenges associated with the use of generative AI in AAC devices. 
We found that generative AI created communication options align well with those created by SLPs and AAC researchers. However, we also discovered challenges. Specifically, SLPs draw on a deep understanding of the individual contexts and backgrounds of their clients to provide tailored support, a level of personalization that generative AI currently lacks. Additionally, it is unclear how harmful developmentally inappropriate communication options that may be generated would be on their language development.

\section{Study 1: Comparing Human- and LMM-Generated Communication Options}

We created a corpus of COs for multiple scenarios sourced from people and LMMs. We collected the human-created COs by surveying SLPs and AAC researchers. We created the LMM-generated COs using OpenAI's GPT-4V\footnote{\href{https://openai.com/gpt-4}{https://openai.com/gpt-4}}, the most popular LMM at the time this work was done. We then conducted a comparison between the human- and LMM-generated COs using a combination of deductive coding, part-of-speech (POS) analysis, and expert analysis.

\subsection{Communication Option Collection} 
%Our team consists of SLPs and AAC researchers and we relied on their network to disseminate the surveys via email. This recruitment resulted in 13 survey responses from SLPs and AAC researchers. 
We collected COs from SLPs and AAC researchers (N=13) via a survey.
%Of our participants, 8 were licensed SLPs. 
The range of professional or clinical experience for our participants was between 1 and 25 years ($\bar{x} =9.3$). The majority (7/13) of our participants were frequent users or researchers of VSDs.
Participants created COs for three contexts: playing, reading a storybook, and retelling a past activity. These contexts were selected because they are common use cases for VSDs~\cite{laubscher2019effect, chapin_effects_2022, bhana_supporting_2020}. We selected six different images, two for each context. See Figure~\ref{fig:sample-vsd-w-comm-options} for a sample image shown to participants. For each context, participants created COs for one image per context which was randomly chosen between the two image variants in order to create a variety of COs for future comparison. 
%To contextualize the task for participants, we used vignettes written by a member of our research team who is an experienced AAC researcher and licensed SLP. 
Vignettes, which are common in AAC research~\cite{holyfield_effect_2019}, were provided to participants to provide context about how the VSD would be used and the linguistic abilities of the fictional child using the VSD. 
%Vignettes are commonly used in AAC research to describe AAC users’ language capabilities and needs~\cite{holyfield_effect_2019}. 
Participants were provided with two different vignettes for different communication stages for each context: children working on building engagement in interactions and the emergence of words  (i.e., pre-linguistic) and children focused on beginning to combine words (i.e., multiword). We refer to these communication stages as pre-linguistic and multiword, respectively. A sample vignette for the playing context for each of the communication stage ((A) pre-linguistic and (B) multiword) is included below: 
%\begin{quote}
``You took this photo to create a VSD for the child in the yellow shirt in this photo. Please write out the COs you would program for them if you were focused on [(A) building engagement in interactions and the emergence of words (B) beginning to combine words].''
%\end{quote}
%in Table~\ref{tab:sample-vignette}.
%(see Appendix~\ref{section:all-vignettes} for all vignettes used in our survey).

% \begin{table}[]
% \centering
% \begin{tabular}{ll}
% \toprule
% \textbf{Playing Vignettes} & \textbf{} \\
% \midrule
% Pre-Linguistic & \begin{tabular}[c]{@{}l@{}}You took this photo to create a VSD for the child in the yellow shirt in this photo. \\ Please write out the COs you would program for them if you \\ were focused on building engagement in interactions and the emergence of words.\end{tabular} \\
% \midrule
% Multiword & \begin{tabular}[c]{@{}l@{}}You took this photo to create a VSD for the child in the yellow shirt in this photo. \\ Please write out the communication options you would program for them \\ if you were focused on beginning to combine words.\end{tabular}\\
% \bottomrule
% \end{tabular}
% \caption{A sample vignette for the playing context for both communication stages (pre-linguistic and multiword).}
% \label{tab:sample-vignette}
% \end{table}

%\subsection{Generating Communication Options with Generative AI}
%For every context, image variant, and communication stage (pre-linguistic and multiword), we prompted GPT-4V to generate COs. 
We prompted GPT-4V to generate COs for the same contexts as our human participants.
We generated an equal number of sets of COs as we obtained from our survey participants. Our prompt to the model contained instructions for generating COs, including the communication stage and a vignette similar to what the survey participants saw. An example prompt (using the image in Figure~\ref{fig:sample-vsd-w-comm-options}) for the pre-linguistic context is provided here:
%\begin{quote}
“You're an assistant to generate vocabulary for pre-linguistic communicators on the autism spectrum who use AAC devices in the form of visual screen displays. This photo was taken to be used in a visual screen display for the child in the yellow shirt in the picture. Please write out the most contextually relevant communication options you would program for them if you were focused on building engagement in interactions and the emergence of words.”
%\end{quote}
%This format of prompt resulted in lists of communication options with on average 20.1 different options. 
%An example of a full response from the LMM can be found in Appendix~\ref{section:full-gpt-response}. 
%In order to be able to compare the LMM-generated communication options to the human-created communication options, we used a process known as prompt chaining~\cite{tongshuang2022ai} which has the model generate an initial output which becomes the input to a second prompt. We used this technique to first generate communication options and then to prompt the LMM to select the most relevant communication options from the original list. The prompt-chaining prompt is included below:
    %\begin{quote}
    %“Using the communication options you generated, please identify the five most relevant communication options.”
    %\end{quote}
%The number five was chosen because it was the closest whole number to the average number of communication options created by expert participants in our survey ($\bar{x} = 4.6 \pm 2.5$).

%\subsubsection{Part-of-Speech Analysis}
\subsection{Communication Option Analysis}
First, we conducted a POS analysis to determine how the content and structure of the COs aligned across the human- and LMM-generated COs. In order to ensure no parts-of-speech were missed, we split every CO into single words.
%\subsubsection{Deductive Coding}
Additionally, to evaluate differences and similarity in the focus of generated COs, we conducted a deductive coding process for both the human- and LMM-generated COs. We used the `'Four Functions of Communication'' framework~\cite{light1988interaction} as our coding scheme: expressing wants or needs (intended to make requests), information transfer (intended to share information with others), social closeness (intended to develop or maintain relationships), and social etiquette (intended to convey polite terms (e.g., ``thank you'')).
%\begin{enumerate}
%    \item Expressing Wants or Needs - communication intended to make requests
%    \item Information Transfer - communication meant to share information with others
%    \item Social Closeness - communication meant to develop or maintain relationships
%    \item Social Etiquette - communication meant to convey polite terms (e.g., “thank you”, “please”, “hello”)
%\end{enumerate}
We also included an ``Other'' category to handle COs which did not clearly align with these four functions. Two researchers performed the coding and inter-rater reliability was computed.
%in accordance with best practices for qualitative research~\cite{mcdonald2019reliability}. 
%We computed the inter-rater reliability using Cohen’s Kappa since we had two raters and categorical codes~\cite{cohen1960coefficient}. 
The inter-rater reliability score was 0.65 indicating substantial agreement between raters~\cite{landis1977application}.
%There is an important limitation of this coding scheme which warrants discussion. The communication options, both human and LMM, were analyzed without knowing the intention of how the communication option would be used. For example, if the communication option was “ball” it is impossible to know if the function of the communication is to express wants and needs (e.g., “throw me the ball”) or for information transfer (e.g., “I see a ball”).

%\subsubsection{Quality Comparison of Communication Options}
After we compared the content and focus of the COs, we conducted a follow-up survey with a subset of experts (N=5) (i.e., more than 5 years of experience and extensive experience using and configuring VSDs) participants from our first survey. Each participant was shown COs for the same contexts as our first survey. Participants did not rate the COs they previously created.
%To ensure that participants did not see the communication options they created in the previous survey, they were shown the other image for each context. For example, if a participant created communication options for Image A for the ``reading a storybook'' context, they would be shown communication options generated by humans and the LMM for Image B in this portion of the study. 
Participants were not aware that any options had been LMM-generated. 
%For each context and communication stage, 
Participants rated LMM- and human-generated COs on a scale from 1 to 5 (1 being the worst). Each participant rated between 68 and 80 sets of COs for a total of 364 ratings.

% \subsubsection{Confidence Analysis}
% For the human-created communication options, we used the participants’ self-reported confidence levels to compute the average confidence level of experts versus non-experts, as well as the confidence levels of experts and non-experts across contexts. Confidence levels were reported on a five-point Likert-scale (Not at all Confident, Not so Confident, Somewhat Confident, Very Confident, Extremely Confident). Confidence was analyzed by comparing the average confidence of experts and non-experts broadly and across all contexts they created communication options for.

% \begin{table}[]
% \centering
% \begin{tabular}{ll}
% \toprule
% \textbf{Source} & \textbf{Communication Options} \\
% \midrule
% Human (Expert) - P4 & ``children's names, build, vroom'' \\
% \midrule
% Human (Non-Expert) - P5 & ``3...2...1... go!, more, let's play together'' \\
% \midrule
% GPT-4V & ``play, block, more?, happy, my turn, your turn'' \\
% \bottomrule
% \end{tabular}
% \caption{Sample communication options created by people and from GPT-4V for the Playing context shown in Figure~\ref{fig:sample-img}.}
% \label{tab:sample-comm}
% \end{table}

\subsection{Results}
In total, human participants created 306 COs across all contexts and the LMM generated 379 (see Figure~\ref{fig:sample-vsd-w-comm-options} for examples). From the results of our POS analysis and deductive coding show that the content and focus of human- and LMM-generated COs are very similar.
Table~\ref{tab:pos-freq} shows the POS frequencies for the COs created by survey participants and the LMM. There is generally alignment in the structure and content of the COs. Nouns and verbs are the most commonly used parts-of-speech for both people and the LMM. 

% \begin{table}[]
% \centering
% \begin{tabular}{ccc}
% \toprule
% \textbf{Part of Speech} & \textbf{Human} & \textbf{LMM} \\ 
% \midrule
% Adjectives & 7.60\% & 10.10\% \\ 
% Adverbs & 5.00\% & 8.90\% \\
% Interjections & 1.70\% & 2.30\% \\
% Nouns & 39.2\% & 33.8\% \\
% Particles/Determiners/Conjunctions & 6.40\% & 1.20\% \\
% Prepositions & 5.20\% & 5.10\% \\
% Pronouns & 6.70\% & 6.90\% \\
% Sound Effects & 2.40\% & 0\% \\
% Verbs & 25.8\% & 31.7\% \\
% \bottomrule
% \end{tabular}
% \caption{Part-of-speech frequencies for all human-created and LMM-generated communication options.}
% \label{tab:pos-freq}
% \end{table}

\begin{table}
\small
\parbox{.45\linewidth}{
\centering
\begin{tabular}{ccc}
\toprule
\textbf{Part of Speech} & \textbf{Human} & \textbf{LMM} \\ 
\midrule
Adjectives & 7.60\% & 10.10\% \\ 
Adverbs & 5.00\% & 8.90\% \\
Interjections & 1.70\% & 2.30\% \\
Nouns & 39.2\% & 33.8\% \\
Particles/Determiners/Conjunctions & 6.40\% & 1.20\% \\
Prepositions & 5.20\% & 5.10\% \\
Pronouns & 6.70\% & 6.90\% \\
%Sound Effects & 2.40\% & 1.0\% \\
Verbs & 25.8\% & 31.7\% \\
\bottomrule
\end{tabular}
\caption{The distribution of part-of-speech for COs.}
\label{tab:pos-freq}
}
\hfill
\parbox{.45\linewidth}{
\centering
\begin{tabular}{ccc}
\toprule
\textbf{Code} & \textbf{Human} & \textbf{LMM} \\
\midrule
Expressing Wants and Needs & 16.7\% & 16.9\% \\
Information Transfer & 69.9\% & 79.4\% \\
Social Closeness & 6.2\% & 0.8\% \\ 
Social Etiquette & 2.0\% & 1.8\% \\
Other & 5.2\% & 2.3\% \\ 
\bottomrule
\end{tabular}
\caption{The results of the deductive coding using Light's~\cite{light1988interaction} Four Functions of Communication framework.}
\label{tab:code-freq}
}
\end{table}

% \begin{table}[]
% \centering
% \begin{tabular}{ccc}
% \toprule
% \textbf{Code} & \textbf{Human} & \textbf{LMM} \\
% \midrule
% Expressing Wants and Needs & 27.5\% & 48.6\% \\
% Information Transfer & 4.25\% & 2.11\% \\
% Social Closeness & 10.5\% & 6.3\% \\ 
% Social Etiquette & 55.6\% & 42.2\% \\
% Other & 2.29\% & 0.79\% \\ 
% \bottomrule
% \end{tabular}
% \caption{Results of performing deductive coding using Light's~\cite{light1988interaction} four functions of communication.}
% \label{tab:code-freq}
% \end{table}

%\subsubsection{Deductive Coding}
In addition to content and structure, the focus of COs was fairly similar for human- and LMM-generated COs (see Table~\ref{tab:code-freq}).
%In terms of the focus of COs generated, Table~\ref{tab:code-freq} shows the results of performing deductive coding using Light’s~\cite{light1988interaction} framework. 
People focused primarily on information transfer and expressing wants and needs, which is congruent with existing AAC research~\cite{ganz2015aac, holyfield2017systematic, light2002there}. Similarly, the LMM also focused mostly on information transfer and the expression of wants and needs. However, the LMM generated very few COs for social closeness. Additionally, participants generated 2.2 times as many COs which belonged to the other category which were commonly sound effects (e.g., “vroom”, “bawk bawk”). We observed a similar focus on social etiquette between the human- and LMM-generated COs.

\begin{figure*}
\centering
\includegraphics[width=0.7\linewidth]{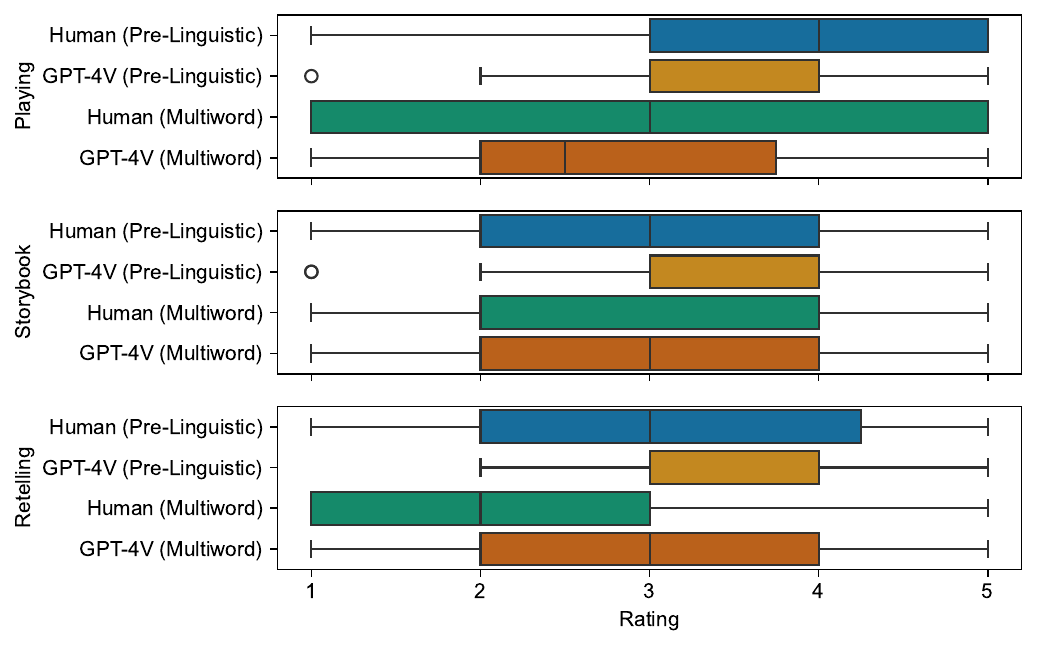}
\caption{A comparison of the experts' ratings of COs generated by human participants and GPT-4V. Human-generated COs were preferred for the Playing context; however, for the Storybook and Retelling contexts, LMM-generated COs were preferred.}
\label{fig:rating-comp}
\Description[A comparison of ratings of human- and LMM-generated communication options.]{In this image, a comparison of expert ratings of human- and LMM-generated communication options are presented via three boxplots. There are boxplots for each context: Playing, Reading a Storybook, and Retelling a Past Activity. Each context has two linguistic abilities: Pre-Linguistic and Multiword. In the Playing context, VSD experts preferred the human-generated communication options. In the other two contexts, Reading a Storybook and Retelling a Past Activity, the experts preferred the LMM-generated communication options.}
\end{figure*}

%\subsubsection{Ratings of Communication Options}
Based on our expert evaluation, LMMs can generate COs of similar, if not better quality than humans (see Figure~\ref{fig:rating-comp}).
%Beyond communication content and focus, Figure~\ref{fig:rating-comp} contains a comparison of the ratings of the human- and LMM-generated COs across all contexts and communication stages. 
For the playing context, the human-generated COs were generally preferred over the LMM-generated COs. For the reading of a storybook and retelling of a past activity, however, the LMM-generated COs were generally preferred over the human-generated COs. While there are differences in the average ratings, the human- and LMM-generated COs perform similarly across all contexts.

\section{Study 2: Semi-Structured Interviews with Clinicians and AAC Researchers}

%To elicit the preferences of SLPs and researchers about the use and ethics of generative AI for automated programming of VSDs, 
We conducted semi-structured interviews (N=5) with SLPs and AAC researchers to begin to understand potential benefits and issues with using generative AI for automation of VSDs. 
%Each interview lasted between 15 and 20 minutes.
%We specifically recruited AAC researchers in order to obtain more insights about potential future research directions in addition to eliciting their perceptions on the use of generative AI for VSDs. 
%The participant backgrounds included university faculty, postdoctoral researchers, or doctoral students in communication sciences and disorders 
Participants had between 5 and 15 years of experience ($\bar{x} = 9$) using VSDs clinically or in intervention research and all participants have published articles on VSDs, vocabulary selection for beginning communicators, or JIT programming for AAC devices.
%Once the interviews has been conducted and transcribed, 
One researcher on our team performed a thematic analysis~\cite{braun2006using} on the interview transcripts by reviewing participants' responses, coding participants' insights, and identifying themes.

%\subsection{Methodology}
%\subsubsection{Participants}
%We conducted 5 semi-structured interviews with AAC researchers
%this smaller sample size is appropriate in this context due to our limited access to this population and prior work has shown that high-quality insights can be obtained even with smaller sample sizes~\cite{ahsen2022designing, allen2007design}. 
%\subsubsection{Interview Protocol}
%We used a semi-structured interview approach to facilitate asking follow-up questions for additional details and perspective~\cite{magaldi2020semi}. We asked participants what kinds of features and functionality would make automated just-in-time VSDs more helpful for VSD users and communication partners. As well as what potential benefits and issues they envision with using generative AI for automated JIT programming and ways they suggest mitigating those risks. Once the interviews had been conducted and transcribed,
%\subsubsection{Thematic Analysis} 
%one researcher on our team performed a thematic analysis~\cite{braun2006using} on the interview transcripts, by reviewing participants’ responses, coding participants’ insights, and identifying themes.
%The structure of our semi-structured interview allowed the researcher to determine preliminary themes of participants’ insights using the following coding scheme: benefits of generative AI in AAC and potential risks of generative AI in AAC. Following these initial themes, sub-themes were identified and analyzed.

\subsection{Results} 
Multiple themes emerged through our interviews about the potential benefits of incorporating generative AI in AAC devices for automated programming, including the reduction of effort required for JIT programming by SLPs and improved accessibility of VSDs and other AAC for untrained communication partners. Two primary themes emerged regarding participants' concerns about potential harms of LMM-generated COs: lack of personalization of LMM-generated COs and developmentally inappropriate LMM-generated COs.
Four participants (P1, P2, P3, P4) expressed concerns about how AI-generated COs lack the context that exists between communication partners and AAC users, and as such, will likely miss out on personally relevant COs, especially regarding culturally relevant or family-oriented COs. 
%Participant 3 said, ``There's a dyadic relationship with professionals or parents and a child and the child can directly point at a picture and show them what they want. So if a child's really interested in one character, and they can point to it, the partner can add it and they're guaranteed to have that vocabulary. Whereas, if the VSD is just reliant on the AI, is it going to pick up what the child wants to say or not?”
%Participant 4 brought up the importance of the background knowledge that SLPs have about a client that an AI system isn’t privy to, saying, “There are initial vocabularies that are universal across a lot of individuals, but you also have very specific vocabulary that are important to each client.”
% \begin{quote}
% \textit{“There are initial vocabularies that are universal across a lot of individuals, but you also have very specific vocabulary that are important to each client.” }
% \end{quote}
%\textbf{Participants were also concerned about developmentally inappropriate LMM-generated COs.} 
Four participants  (P1, P2, P3, P5) were concerned about the potential risks of dynamically generated communications. Specifically, they were concerned about the harm COs which are not developmentally appropriate could pose to advancement of communication skills.

\section{Discussion \& Future Work}

%In this paper, we presented the results of three studies. The first two studies compared the relevance and focus of human-created and LMM-generated communication options for three contexts to identify where these two sets of communication options aligned and misaligned. We also presented the results from five semi-structured interviews with VSD experts to gain further insights into the potential application of AI in this context and the ethical and practical implications.
%With the highly performant abilities of generative AI, we can begin to grapple with the question of how using AI-generated COs in AAC devices would fundamentally change the communication process of existing JIT configured systems. 

%Our findings indicate that the content, focus, and quality of LMM-generated COs were generally comparable to those created by humans. However, human-generated COs often included more sound effects and language aimed at social engagement. Furthermore, human-generated COs provided additional contextual information about users that went beyond what was present in any given image. For example, SLPs wanted to include the names of the kids in the photographs. 

Our findings indicate that the content, focus, and quality of LMM-generated COs were generally comparable to those created by humans. However, human-generated COs often included more sound effects and language aimed at social engagement. This difference represents a minor departure from best practices~\cite{hoff2006social, vygotsky2012thought, light2002there, fenson1994variability, nelson1973structure}. 

Furthermore, our interview study identified concerns participants had about whether the COs would be developmentally appropriate or personally relevant for users. This highlights a key value that SLPs bring to the usage of VSDs---detailed knowledge about the user's interests and their linguistic ability. 
Personalization is a key aspect of early language development~\cite{frick2024examining, laubscher2020core, light2021personalized}, and, currently, generative AI does not provide the personalization necessary for the effective use of VSDs for early language development. Future work could explore the development of personalized user models to address this issue; however, 
%Returning to RQ1, we find that the quality of responses was roughly equivalent between people and LMMs but this only generalizes to this narrow context where the identity of the individual is not fully taken into account. 
%More work is needed to understand what the implications might be for language learning. 
%It is possible that these language options might be useful in the moment, but potentially harmful for long-term language development. 
%Future work should more deeply investigate how developmentally appropriate COs given by generative AI are and the trade-offs which happen with more contextually-relevant COs rather than more personalized COs. Additionally, 
the value SLPs and communication partners bring to interactions with VSDs cannot and should not be replaced. Future work should leverage communication partners' insights to develop a better understanding of how they can monitor and edit generated COs to ensure their relevance and appropriateness. 
LMMs have been observed to produce harmful output which contains negative stereotypes and biases~\cite{navigli2023biases, armstrong2024silicone, ding2023fluid, kotek2023gender}. While we did not observe any occurrences of this in our research, it is critical that future work prioritizes identifying mitigation strategies to prevent these stereotypes and biases from being present in any AAC devices relying on LMM output.
%as we did not include VSDs users in the current study due to the methodology in this study requiring participants to use language children who are learning words through VSDs have not yet developed~\cite{frauenberger2012challenges}, 
Lastly, future work should also incorporate appropriate methods to reduce linguistic demands ~\cite{frauenberger2012challenges} allowing for VSD end users to participate in the design process.

\begin{acks}
We are grateful to Abi Roper for her guidance and expertise which helped refine the framing of this poster. This research was partially funded by the Temple University College of Science and Technology Research Scholars Program and Convergence Accelerator Grant (National Science Foundation grant number ITE-2236352).
\end{acks}

\bibliographystyle{ACM-Reference-Format}
\bibliography{sample-base}

\appendix
\section{LMM Prompting Methodology}

In order to collect the LMM-generated communication options, we used OpenAI's GPT-4V model, specfically the \texttt{gpt-vision-preview} model. All responses were collected in April 2024. We used two types of prompts. The first prompt type was intended to generate communication options for communicators working on building engagement in interactions and the emergence of words (pre-linguistic): 
\begin{quote}
  \textit{ ``You're an assistant to generate vocabulary for pre-linguistic communicators on the autism spectrum who use AAC devices in the form of visual scene displays. This photo was taken to be used in a visual scene display for [a child pictured in the image]. Please write out the most contextually relevant communication options you would program for them if you were focused on building engagement in interactions and the emergence of words.''} 
\end{quote}
The second prompt was intended to generate communication options for communicators focused on beginning to combine words (multiword): 
\begin{quote}
    \textit{``You're an assistant to generate vocabulary for multiword communicators on the autism spectrum who use AAC devices in the form of visual scene displays. This photo was taken to be used in a visual screen display for [a child pictured in the image]. Please write out the communication options you would program for them if you were focused on beginning to combine words.''}
\end{quote}
These prompts were chosen because they included necessary contextual information about the end user of the visual scene display (e.g., diagnosis, linguistic ability), and closely mirrored the vignette which was provided to our human participants. 

The above prompts results in lists of communication options with on average 20 options. In addition to using the above prompts, we used a technique called prompt chaining~\cite{tongshuang2022ai} to narrow the lists of communication options to the 5 options deemed most relevant by the model. The chaining prompt we used was: 
\begin{quote}
    \textit{``Using the communication options you generated, please identify the five most relevant communication options.''}
\end{quote}

We acknowledge that there might exist prompts which result in more contextually relevant results. The key limitation of LMMs generating vocabulary for AAC devices, which we identified in this work, is the lack of personalization of the responses, regardless of the quality of the contextually relevant communication options.
\end{document}